\documentstyle[12pt]{article}
\begin{document}
\title{Hidden Symmetry and S-duality in N=4 D=4 Super Yang-Mills Theory}
\author{I.Ya.Aref'eva \\ Steklov Mathematical
Institute, \\ Russian Academy of Science, Moscow , Russia
\thanks{ E-mail :
arefeva@arevol.mian.su},}
\date {$~$}
\maketitle
\begin {abstract}

In this talk the Schwarz hypothesis that the duality symmetries should
be pieces of the hidden gauge symmetry in a string theory is discussed.
Using auxiliary linear system special dual transformations for $N=4$ SYM
generalizing the Schwarz dual transformations for a principal chiral model
are constructed. These transformations are related with  a continuous group
of hidden symmetry of a new model involving more fields as compare
with $N=4$ SYM. We conjecture that the $Z_{2}$ discrete subgroup of this
hidden symmetry group has  a stable set of N=4 YM fields and  transforms
a self-dual configuration to an anti-self-dual and via versa.
\end{abstract}
\section{Introduction}
\setcounter{equation}{0}
In the recent years, the idea that string theory will
exhibit an enormous gauge symmetry group has got many supports.
There are the following indications supporting this idea.

$\bullet $ More then ten years ego it was realized that
the theories which now are considered as describing a low energy limit of
string theories
do posses some hidden symmetries. Namely,
there are global continuous non-compact symmetries  in classical
supergravity theories. A non-compact $SL(2,R)$ symmetry group was found
in $N=4, ~D=4$ supergravity by Cremer, Ferrara and Scherk
\cite {CFS} and $E_{7}$ symmetry was found in $N=8$, $D=4$
supergravity by Cremer and Julia \cite {Julia} (about other examples see
\cite{ssrg}).
More  yearly Geroch \cite {gr}  has found special solutions of Einstein's
equations  that were invariant under an infinite set of transformations
(about further  developments see   \cite {grf} and about applications to
strings see \cite {str}).

 $\bullet $ The existence of  the Lax representation for $N=4$ $D=4$ and
$N=1$, $D=10$ super Yang-Mills theories \cite {Vol} was associated,
due to wide experience in  two-dimensional integrable models \cite {LD},
with some hidden symmetry.
For these models there are infinite number of conserved currents
\cite {Dev,AV} generalized the Lucher-Pohlmeyer-Brezin et al.
currents \cite{LP,BIZZ} of two-dimensional integrable models.
In the two-dimensional cases current conservations provide
enough restrictions in order to solve the theory exactly.

$\bullet $  A search for a symmetry in string theory  was one
of motivation of the Witten construction of the string field theory
\cite{Witten}.

$\bullet $ The idea of enormous symmetry in string theory was
advocated by Gross and Mende in the context of high energy scattering
of strings \cite{GM}.

$\bullet $ This idea got essentially more support after discovering
the duality  symmetry in string theory. Few years ego, Font, Inanez, Lust
and Quevedo \cite{Lust} proposed that an $SL(2,Z)$
discrete subgroup
of SL(2,R) should be an exact symmetry of the heterotic string toroidally
compactifies to four dimensions (see also  \cite {Sch2,Sen}).
This duality symmetry in a special case implies an electric-magnetic
duality and is intrinsically non-perturbative
and therefore cannot be test within perturbation theory.
Initial evidence in favour of s-duality for N=4
SYM was provided by the exact agreement of masses of particles
and solitons with those predicted by the
$Z_{2}$ symmetry of Montenen and Olive.
Tests of full $SL(2,Z)$ s-duality have been formulated and verified
in \cite{Gir,VW}. Girardello et al  \cite {Gir} and
Vafa and Witten   \cite {VW} have examined the
implications of s-duality for N=4 SYM,
and Seiberg and Witten  \cite {SW} have examined
N=2 SYM.

$\bullet $ There is the Schwarz conjecture \cite{Sch} that the duality
symmetries should be pieces
of the hidden gauge symmetry in a string theory.
The string duality symmetries are infinite discrete groups.
The conjecture is that they are the discrete subgroups of
gauge symmetries of string theory.
Note that symmetry groups appears in the classical supergravity are
continuous ones. At first sight there is a discrepancy  between
continuous symmetry of classical theories and discreteness of the group of
duality transformations.
One can expect that the restrictions to discrete groups are due
to quantum corrections or sting effects.
In particular, Hull and Townsend  \cite {HuTo} have shown that $E_{7}$ symmetry
of N=8,
D=4 supergravity found in \cite{Julia} is broken by quantum
effects to a discrete subgroup, $E_{7}(Z)$, which contains both the
T-duality group $O(6,6,Z)$ and the S-duality group  $SL(2,Z)$.

But another point of view is also admissible. Apparently,
there is no raison to expect that  classical and quantum properties
of N=4, D=4 , which describes a low  energy sector of string theory after
trivial reduction, are strongly different.
The validity of this opinion  relies on the existence of non-renormalization
theorems. In the example considered by Hull and Townsend
non-renormalization theorem is not valued.
If so, it is not worth to try to find an continuous hidden symmetry group
for N=4 SYM.
Note that in spite of existence of conserved currents the hidden symmetry for
N=4, D=4
super Yang-Mills is still not found. We will see on the model example that
the existence of the continuous duality group
can be depended on the boundary conditions of the model under consideration,
or on a coefficient in front of an topological  term.

We are going to start from a consideration
the Schwarz dual transformation for the D=2  principal chiral model
with anomaly term.
Note  that since duality
is an intrinsically non-perturbative property
it is appropriate in this context
to use our knowledge concerning
D=2 exact integrable models \cite{LD}.
We will see that for all possible $\theta$
except $\theta =1$, it is possible to do dual transformation,
parametrized by some parameter $\lambda$,
 $-(1+\theta)<\lambda < 1+\theta $ getting
 a new $\theta $, say  $\theta ^{\prime}$
 according the formula
\begin {equation}
                                      \label {T'}
\theta^{\prime }
=\frac{\lambda -\theta(1+\theta)}
{\lambda \theta-(1+\theta)}.
\end{equation}

We see from this formula that $\theta =\pm 1$ are the stable points
of this  transformation.
In these special points one can make another dual transformation which makes
 a change
\begin {equation}
                                      \label {T}
\theta^{\prime }=-\theta
\end{equation}
This gives  an example then the
continuous group in special values of parameters of
the model under consideration shrink to
discrete $Z_{2}$  group.

We expect the similar picture for $N=4$, $D=4$ being embedded
in a  more large model. In particular,
there is so called N=4, D=4 B-model which as compare with
N=4 SYM contains one extra dynamical field $B$. For $B=0$
this model coincides with SYM. This model is non-relativistic one,
but it is supersymmetric and gauge invariant.
This model possesses a
hidden symmetry. Applying  infinitezimally
these symmetry transformations
to N=4 SYM we get
non-zero value for field B and we abandon mass-shell of N=4 SYM
and occur on mass-shell of B-model.
But it may happen that after some global transformation  we
will come back to B=0. In this case one can say that
continuous group
of the large model (B-model in our case) is reduced to a discrete subgroup
of more restricted model, i.e. model which has more constrains (N=4 SYM
in our case).
It is tempting to expect  that a transformation which
transforms a self-dual configuration to an anti-self-dual one
for SYM model can be considered as a global hidden symmetry transformation
for B-model.

\section   {Dual Transformations for Principal Chiral Models}
 \setcounter{equation}{0}
An equation of motion for a principal chiral model (PCM)
with an anomaly (a WZNW term) has the form
\begin{equation}
                                          \label{cse}
\partial _{\mu}A_{\mu}+\theta \epsilon _{\mu \nu}\partial _{\mu}A_{\nu}=0,
\end{equation}
where
\begin {equation} 
                                                          \label {dg}
A_{\mu}=g^{-1}\partial _{\mu}g,
\end   {equation} 
so $A_{\mu}$ satisfies also the zero-curvature condition
\begin {equation} 
                                                          \label {bi}
F_{\mu \nu}=\partial _{\mu}A_{\nu}-\partial _{\nu}A_{\mu}+
[A_{\mu},A_{\nu}]
\end   {equation} 
It is convenient to introduce light-cone coordinates
\begin {equation} 
                                                          \label {lc}
x^{\pm}=x^{0}\pm x^{1},
\partial _{\pm}=\frac{1}{2}(\partial _{0}\pm \partial _{1}),
\end   {equation} 
so that the equation of motion takes the form
\begin{equation}
\label{se}
\partial _{+}A_{-}+\kappa \partial _{-}A_{+}=0.
\end{equation}

Let us consider the Schwarz dual transformation  \cite {Sch}
\begin{equation}
\label{A''}
A_{\mu} ~\to ~A^{\prime  }_{\mu}=
g^{\prime   -1}\partial _{\mu}g^{\prime  },
\end{equation}
where
\begin{equation}
\label{g''}
g^{\prime  }=g(\theta ,x)\Psi (\theta , \lambda),
\end{equation}
$g(\theta ,x)$ is a solution of (\ref{cse}) and
$\Psi (\theta, \lambda)$ is a solution of the following
linear system of equations
\begin{equation}
\label{p'}
(\partial _{+}-
\frac{\lambda (1-\theta )}{1+\theta -
\lambda}A_{+}(\theta ,x))\Psi (\theta , \lambda)
=0,
\end{equation}

\begin{equation}
\label{p''}
(\partial _{-}+\frac{\lambda (1+\theta )}{1+\theta +\lambda}
A_{-}(\theta ,x))\Psi (\theta , \lambda)
=0,
\end{equation}
                    \label{lr}
\begin{equation}
-(1+\theta )<\lambda < 1+\theta  .
\end{equation}

Simple calculations show that after the dual transformation of the field
$g(c,x)$ we get once again the chiral field with a  new coefficient
in front of the WZNW term.
Namely, one can check that
\begin{equation}
\label{e''}
\partial _{\mu} A^{\prime   }_{\mu}(\theta ,x)+
\theta ^{\prime   }\epsilon _{\mu \nu}\partial _{\mu}
A^{\prime   }_{\nu}(\theta ,x)=0
\end{equation}
with
\begin{equation}
\label{c''}
\theta ^{\prime   }
=\frac{\lambda -\theta (1+\theta )}
{\lambda \theta -(1+\theta )}.
\end{equation}

The function (\ref{c''}) has the following interesting property:
 starting from  $\theta=\pm 1$     after the dual
transformation with arbitrary $\lambda$  we get $\theta^{\prime   }=\pm 1$,
i.e {\it the point} $\theta=\pm 1$ {\it is stable under the dual
transformation.}

As Schwarz pointed out  \cite {Sch}
starting from $\theta=0$ after the dual transformation
with a suitable $\lambda$ one can get any $-1<\theta^{\prime }<1$.
The same is true for arbitrary $\theta\neq \pm 1$. In
particular, starting from
$\theta\neq 1$ after the dual transformation
with a suitable $\lambda$ (which is very closed to the boundary
$\lambda =-(1+\theta)$ ) one can get $\theta^{\prime  }$
which is very closed to  $1$, but never we can get exactly
$\theta^{\prime  }=1$ if $\theta\neq 1$. Formally, to   get exactly
$\theta^{\prime  }=1$
we have to make the dual transformation
with $\lambda =-(1+\theta)$. However just this point is
a singular point for linear equations (\ref{p''}).
So we can say that in the points $\theta=\pm 1$
there are isolated  "traps".

The same picture takes place for a  superchiral field.
Equation of motion for the superchiral field with the anomaly (the WZNW term)
has the form
\begin{equation}
\label{e}
D_{1}A_{2}-\kappa D_{2}A_{1}=0,
\end{equation}
where $A_{i}$ is defined as
\begin{equation}
\label{c}
A_{i}=G^{-1}(y)D_{i}G(y).
\end{equation}
Here  $G(y)$ is a superfield
$$G(y)=\exp \sum _{a}T_{a}\phi _{a}(x_{\mu}, \theta_{1},
\theta _{2}),~~y_{A}=({\mu},\theta _{1}, \theta _{2}), $$
\begin{equation}
\label{D}
D_{1}=\frac{\partial }{\partial \theta _{2}}-i\theta _{2}
\frac{\partial}{\partial x^{+}}, ~~
D_{2}=-\frac{\partial }{\partial \theta _{1}}+i\theta _{1}
\frac{\partial}{\partial x^{-}}, ~~
\end{equation}
$x^{+ }$ and $x^{-}$
are light-cone coordinates (\ref {lc}).
Supercurrent $A_{i}$ also satisfies the zero-curvature identity
\begin{equation}
\label{zc}
D_{1}A_{2}+D_{2}A_{1}+\{A_{1}, A_{2} \}=0.
\end{equation}
Compatibility conditions for the following linear system  \cite {Volsc}
\begin{equation}
\label{l}
D_{1}\Psi (\lambda)=\frac {\lambda \kappa}{1-\lambda \frac{1+\kappa}{2}}
A_{1}\Psi
\end{equation}
\begin{equation}
\label{ll}
D_{2}\Psi (\lambda)=-\frac {\lambda }{1+\lambda \frac{1+\kappa}{2}}
A_{2}\Psi .
\end{equation}
are just equations (\ref {e}) and (\ref{zc}). To recall that $G (y)$
is a solution of equation (\ref{e}) with some  $\kappa$
we write
$G_{\kappa }(y)$ and $\Psi _{\kappa }(\lambda)$ .

Now let do an analog of the transformation (\ref {g''}). Multiplying
$G_{\kappa}(y)$  on $ \Psi _{\kappa}(\lambda)$
\begin{equation}
\label{sd'}
G^{\prime}=G_{\kappa}\Psi _{\kappa}(\lambda)
\end{equation}
one can check that
\begin{equation}
\label{A1}
A^{\prime}_{i}=G^{\prime   -1}(y)D_{i}G^{\prime  }(y)
\end{equation}
satisfies the following equation
\begin{equation}
\label{s''}
D_{1}A^{\prime  }_{2}-\kappa ^{\prime  }
D_{2}A^{\prime  }_{1}=0,
\end{equation}
where
\begin{equation}
\label{k''}
\kappa ^{\prime  }=\kappa \frac{1+\lambda \frac{1+\kappa}{2}}{1-
\lambda \frac{1+\kappa}{2}}
\end{equation}

And once again we have  stable points of the dual transformation.
They are $\kappa =0$
and $\kappa =\infty $

\section   {Dual Transformations for N=4 D=4 Super Yang-Mills Theory}
\setcounter{equation}{0}

For N=4 SYM it is possible to do an analog of the Schwarz dual transformation
(\ref {g''}).
Namely, starting form superpotential which is  a solution of
the self-dual equations  one can construct a superpotential which is a
solution of  the  anti-self-dual equations.
This can be consider as an analog of  transformations  which
changes the sign in the front of the WZNW term.
To do this  I'l use  a linear system
of equations which compatibility conditions coincide with
full SYM equation  \cite {Vol}.

Let $C^{4,4N}$ be the complex Minkowski superspace with coordinates

$$y^{A}=(x^{\mu}, \theta ^{\alpha}_{s}, {\bar \theta}^{{\dot \alpha}t}),~
{}~\mu =0,1,2,3;~~ \alpha , {\dot \beta}=1,2;~~s,t=1,...N=4$$

Covariant derivatives have the form
$${\cal D}_{A}=D_{A}+[A_{A}, ~.~],$$
where
$$D_{A}=(\partial _{\mu}, D^{s}_{\alpha},D_{{\dot \beta }t}), ~
D^{s}_{\alpha}= \frac{\partial}{\partial \theta ^{\alpha}_{s}}
+i {\bar \theta }^{{\cdot \beta}s}\partial _{\alpha {\cdot \beta}},~~
D_{{\dot \beta }t}),~~D_{{\dot \beta }t}=-
\frac{\partial} {\partial {\bar \theta} ^{{\dot \beta}t}}
-i  \theta ^{ \alpha }_{t}\partial _{\alpha {\dot \beta}}
\partial _{\alpha {\cdot \beta}}.
$$

Supercurvature $F_{AB}$ is defined as

\begin{equation}
\label{C}
[{\cal D}_{A},{\cal D}_{B}]=T_{AB}^{C}{\cal D}_{C}+iF_{AB},
\end{equation}

where $T_{AB}^{C}$ is a torsion,

$$T_{\alpha, {\dot \beta}t}^{s~~~\mu}=
T_ {{\dot \beta},t \alpha }^{s~~ ~\mu }=-2i\delta ^{s}_{t}\sigma ^{\mu}_
{\alpha {\cdot \beta}}$$

The equations of motion have the form
\begin{equation}
                       \label{F}
F^{(st)}_{\alpha \beta}=0=F_{{\dot \alpha}(s, {\dot \beta}t)}
\end{equation}
\begin{equation}
                       \label{FS}
F^{s}_{\alpha , {\dot \beta}t}=0
\end{equation}
These equations look as the constraint equations,
however it is known that they lead to differential equations in x-space
(put the theory on-shell) \cite {Wit,Gr,Vol},\cite {AVR1}-\cite {LL}.

Let consider a set of linear equations  \cite {Vol}

\begin{equation}
                       \label{Xl}
X^{s}(\lambda)\Psi (\lambda)=0,
\end{equation}
\begin{equation}
                       \label{Yl}
Y_{t}(\lambda)\Psi (\lambda)=0,
\end{equation}
\begin{equation}
                       \label{Zl}
Z(\lambda)\Psi (\lambda)=0,
\end{equation}

where $X^{s}(\lambda), Y_{t}(\lambda)$ and $Z(\lambda)$
are given by the following formulae

$$X^{s}(\lambda)= \nabla ^{s}_{1}+\lambda \nabla ^{s}_{2},$$
$$Y_{t}(\lambda)=\nabla _{{\dot 1}t}+\lambda ^{2}\nabla _{{\dot 2}t},$$
$$Z(\lambda)=\nabla _{1{\dot 1}}+\lambda \nabla _{2{\dot 1}}
+\lambda ^{2} \nabla _{1{\dot 2}}+\lambda^{3} \nabla _{2{\dot 2}}. $$

The integrability conditions of these linear equatioins are

\begin{equation}
                       \label{XX}
 \{X^{s}(\lambda),X^{t}(\lambda)\}=0=
 \{Y_{s}(\lambda),Y_{t}(\lambda)\},
\end{equation}
\begin{equation}
                       \label{XY}
\{X^{s}(\lambda),Y_{t}(\lambda)\}=-2i \delta ^{s}_{t}Z(\lambda)
\end{equation}

One can also write a ``covariant'' linear system
\begin{equation}
                       \label{Xz}
z^{\alpha}\nabla _{\alpha}^{t}\Psi (z,w)=0
\end{equation}
\begin{equation}
                       \label{Yw}
w^{{\dot \alpha}}\nabla _{{\dot \alpha}t}\Psi (z,w)=0
\end{equation}
\begin{equation}
                       \label{Zzw}
z^{\alpha}w^{{\dot \alpha}}\nabla _{\alpha{\dot \alpha}}\Psi (z,w)=0
\end{equation}
One gets (\ref{Xl}), (\ref{Yl}) and (\ref{Zl}) from (\ref{Xz}),
(\ref{Yw}) and (\ref{Zzw}) substituting

\begin{equation}
     \label{lzw}
z^{1}=1,~z^{2}=\lambda, ~~w^{{\dot 1}}=1,~w^{{\dot 2}}=
\lambda ^{2}
\end{equation}

Equations of motion (\ref{F}) and (\ref{FS}) provide the consistency conditions
for the linear system (\ref{Xz})-(\ref{Zzw}).

Equations which  generalize the self-dual equations of the usual Yang-Mills
theory in the superfield notations have
the form  ( \cite {Volsd}, see also  \cite {Sem})
\begin{equation}
                       \label{AF}
F^{st}_{\alpha \beta}=0=F_{{\dot \alpha}(s, {\dot \beta}t)}
\end{equation}
\begin{equation}
                       \label{AFF}
F^{s}_{\alpha , {\dot \beta}t}=0
\end{equation}

For the anti-self-dual case one has
 \begin{equation}
                       \label{AAF}
F^{(st)}_{\alpha \beta}=0=F_{{\dot \alpha}s, {\dot \beta}t}
\end{equation}
\begin{equation}
                       \label{AAFF}
F^{s}_{\alpha , {\dot \beta}t}=0
\end{equation}

There is no symmetrization on the upper $s$ and $t$ in (\ref{AF})
and lower $s$ and $t$ in (\ref{AAF}).

One has the following linear system for the  selfdual super-Yang-Mills
equations
\begin{equation}
                       \label{XA}
\nabla _{\alpha}^{s}\Psi ^{{\cal SD}}(w)=0
\end{equation}
\begin{equation}
                       \label{YA}
w^{{\dot \alpha}}\nabla _{{\dot \alpha}t}\Psi ^{{\cal SD}}(w)=0
\end{equation}
\begin{equation}
                       \label{ZA}
w^{{\dot \alpha}}\nabla _{\alpha{\dot \alpha}}\Psi ^{{\cal SD}}(w)=0
\end{equation}
and the following linear system for the anti-self-dual Super-Yang-Mills
equations

\begin{equation}
                       \label{XAA}
z^{\alpha}\nabla _{\alpha}^{s}\Psi ^{{\cal ASD}}(z)=0
\end{equation}
\begin{equation}
                       \label{YAA}
\nabla _{{\dot \alpha}t}\Psi ^{{\cal ASD}}(z)=0
\end{equation}
\begin{equation}
                       \label{ZAA}
z^{\alpha}\nabla _{\alpha{\dot \alpha}}\Psi ^{{\cal ASD}}(z)=0
\end{equation}

Starting form superpotential
$(A^{{\cal SD}~s}_{~~\alpha},
A^{{\cal SD}}_{{\dot \beta}t}, A^{{\cal SD}}_{\alpha {\dot \beta}})$ being
solution of the self-dual equations (\ref{AF}) and (\ref{AFF})
let construct a solution of anti-self-dual equations
(\ref{AAF}) and (\ref{AAFF}).
To do this let consider the solution of the linear system
of equations (\ref{Xl}), (\ref{Yl}) and (\ref {Zl}) for given superpotential
(compatibility condition takes place since this superconnection solves
(\ref{F}) and (\ref{FS}).

So we have $\Psi (z,w|A_{A}^{{\cal SD}})$. Let now take
\begin{equation}
                       \label{dot}
A_{\alpha}^{\prime s}=A^{{\cal SD}s}_{\alpha},
\end{equation}
\begin{equation}
                       \label{A11}
A^{\prime}_{{\dot 1}t}=A^{{\cal SD}}_{{\dot 1}t}
\end{equation}

\begin{equation}
                       \label{A2}
A^{\prime}_{{\dot 2}t}=A^{{\cal SD}}_{{\dot 2}t}
-[(D_{{\dot 2}}+A^{{\cal SD}}_{{\dot 1}t})\Psi _{1}]\cdot \Psi ^{-1}_{1}
\end{equation}
where
\begin{equation}
                       \label{A22}
\Psi _{1}\equiv \Psi (z,w|A_{A}^{{\cal SD}})\big |_{\displaystyle
{w^{{\dot 1}}=1\atop w^{{\dot 2}}=0}}
\end{equation}

Now one can check that the following equations are fulfilled
\begin{equation}
                       \label{XAA'}
z^{\alpha}\nabla  _{\alpha}^{\prime s}\Psi _{1}(z)=0
\end{equation}
\begin{equation}
                       \label{YAA'}
\nabla ^{\prime}_{{\dot \alpha}t}\Psi _{1}(z)=0
\end{equation}

 where $^{\prime}$ means
$$\nabla '_{{\dot \alpha }t}=D_{{\dot \alpha }t}+A'_{{\dot \alpha}t}$$

Vectors components can be obtained from the spinors components
of the superconnection.
Transformation (\ref {A2}) is a generalization of the Schwarz transformation
(\ref {A''}).

\section {B-model}
\setcounter{equation}{0}

Let us consider a model described by the superconnection $A_{A}(y)$
and an additional superfield $B_{s}(y)$ \cite{AVB}.
These fields satisfy the following
dynamical equations
\begin{equation}
                       \label{FB}
F^{(st)}_{\alpha \beta}=0,~~F_{{\dot 1}s, {\dot 1}t}=0=
F_{{\dot 2}s, {\dot 2}t}
\end{equation}
\begin{equation}
                       \label{BB}
F_{{\dot 1}(t, {\dot 2}s)}+B_{(t}B_{s)}=0,
\end{equation}
\begin{equation}
                       \label{FFB}
{\cal D }_{{\dot \alpha}(t}B_{s)}=0,~~
F^{s}_{1,{\dot 1}t}=0=F^{s}_{2,{\dot 2}t},
\end{equation}
\begin{equation}
                       \label{FFBB}
F^{s}_{2, {\dot 1}t}+{\cal D }^{ s}_{2}B_{t}=0
\end{equation}
This system of equations is non-relativistic one.
But it is invariant under the gauge transformation
\begin{equation}
                       \label{GI}
A_{A}\to K^{-1}A_{A}K +K^{-1}D_{A}K, ~~ B\to K^{-1}B K
\end{equation}
where $K$ is an arbitrary superfield.
If $B=0$ equations (\ref {FB})-(\ref {FFBB}) coincide with
(\ref {F})-(\ref {FS})

Let consider a set of linear systems
\begin{equation}
                       \label{XXl}
{\cal X}^{s}(\lambda)\Psi (\lambda)=0,
\end{equation}
\begin{equation}
                       \label{YYl}
{\cal Y}_{t}(\lambda)\Psi (\lambda)=0,
\end{equation}
\begin{equation}
                       \label{ZZl}
{\cal Z}(\lambda)\Psi (\lambda)=0,
\end{equation}

where ${\cal X}^{s}(\lambda), {\cal Y}_{t}(\lambda)$ and ${\cal Z}(\lambda)$
are given by the following formulae

$${\cal X}^{s}(\lambda)= \nabla ^{s}_{1}+\lambda \nabla ^{s}_{2},$$
$${\cal Y}_{t}(\lambda)=\nabla _{{\dot 1}t}+\lambda B_{t}+
\lambda ^{2}\nabla _{{\dot 2}t},$$
$$Z(\lambda)=\nabla _{1{\dot 1}}+\lambda \nabla _{2{\dot 1}}
+\lambda ^{2} \nabla _{1{\dot 2}}+\lambda^{3} \nabla _{2{\dot 2}}. $$
We see that ${\cal X}$ and ${\cal Z}$ coincide with $X$ and $Z$
respectively, and only there is a difference in the operators
${\cal Z}$ and $Z$.

The integrability conditions of these linear equatioins,
\begin{equation}
                       \label{XXXX}
 \{{\cal X}^{s}(\lambda),{\cal X}^{t}(\lambda)\}=0=
 \{{\cal Y}_{s}(\lambda),{\cal Y}_{t}(\lambda)\},
\end{equation}
\begin{equation}
                       \label{XXYY}
\{{\cal X}^{s}(\lambda),{\cal Y}_{t}(\lambda)\}=
-2i \delta ^{s}_{t}{\cal Z}(\lambda),
\end{equation}
are provided by equations (\ref {FB})-(\ref {FFBB}).

We can write the following hidden symmetry transformation for the
system equations (\ref {FB})-(\ref {FFBB}).
\begin{equation}
                       \label{TA1}
\delta A_{1}^{s}=0, ~~\delta A_{{\dot 1}s}=0, ~~
\delta A_{1{\dot 1}}=0,
\end{equation}
\begin{equation}
                       \label{TA2}
\delta A_{2}^{s}=-\epsilon i {\cal D}_{2}^{s}S(\mu),
\end{equation}
\begin{equation}
                       \label{TA2t}
\delta A_{{\dot 2}t}=-\epsilon i {\cal D}_{{\dot2}t}S(\mu),
\end{equation}
\begin{equation}
                       \label{TA3}
\delta A_{2{\dot 2}}=-\epsilon i {\cal D}_{2{\dot 2}}S(\mu),
\end{equation}
\begin{equation}
                       \label{TA4}
\delta A_{2{\dot 1}}=-\epsilon i \frac{1}{\mu}(\partial _{1{\dot 1}}S(\mu),
\end{equation}
\begin{equation}
                       \label{TA5}
\delta A_{1{\dot 2}}=-\epsilon i ({\cal D} _{1{\dot 2}}S(\mu)+
\mu {\cal D} _{2{\dot 2}}S(\mu)),
\end{equation}
\begin{equation}
                       \label{TB}
\delta B_{t}=i\mu \delta A_{{\dot 2}t}+\epsilon
[B_{t}, S(\mu)],
\end{equation}
where $S(\mu)$ is a solution of the following equations
\begin{equation}
                       \label{S1}
(D_{1}^{s}+\mu {\cal D}_{2}^{s})S(\mu)=0,
\end{equation}

\begin{equation}
                       \label{S2}
(D_{{\dot 1}t}+\mu ^{2}{\cal D}_{{\dot 2}t})S(\mu)=-\mu [B_{t},S(\mu)],
\end{equation}

\begin{equation}
                       \label{S3}
(\partial _{1{\dot 1}}+\mu
{\cal D} _{2{\dot 1}}+\mu ^{2}
{\cal D} _{1{\dot 2}}+\mu ^{3}{\cal D}_{2{\dot2}}
)S(\mu),
\end{equation}
We see that transformations (\ref {TA1}), (\ref {TA2t})
are nothing but an infinitezimal version of (\ref {dot})-(\ref {A2}).

$$~$$
{\bf ACKNOWLEDGMENT}
$$~$$
The author thanks D.Lust, D.Ebert and Institute of Physics of
Humboldt University in Berlin where this work was started for the
kind hospitality.  This work is supported in part by Deutsche
Forschungsgemeinschaft under project DFG 436 RUS 113/29 and in part
by Russian Foundation for Fundamental Research grant 93-011-147.
I am grateful to C.Preitschopf and I.Volovich for stimulating
discussions.

\end{document}